# Crystal and local atomic structure of $Co$-doped $MgFeBO_4$ warwickites

N.V. Kazak[*,1], M.S. Platunov[**,1], Yu.V. Knyazev[2], N.B. Ivanova[2], Y.V. Zubavichus[3], A.A. Veligzhanin[3], A.D. Vasiliev[1,2], L.N. Bezmaternykh[1], O.A. Bayukov[1], A. Arauzo[4], J. Bartolomé[5], K.V. Lamonova[6], and S.G. Ovchinnikov[1,2]

[1]*Kirensky Institute of Physics, 660036 Krasnoyarsk, Russia*

[2]*Siberian Federal University, 660074 Krasnoyarsk, Russia*

[3]*National Research Center "Kurchatov Institute", 123182 Moscow, Russia*

[4]*Servicio de Medidas Físicas, Universidad de Zaragoza, 50009 Zaragoza, Spain.*

[5]*Instituto de Ciencia de Materiales de Aragón, CSIC-Universidad de Zaragoza and Departamento de Física de la Materia Condensada, 50009 Zaragoza, Spain*

[6]*O.O. Galkin Institute for Physics and Engineering, National Academy of Sciences of Ukraine, 83114 Donetsk, Ukraine*

**Abstact** Single crystalline $MgFeBO_4$, $Mg_{0.5}Co_{0.5}FeBO_4$ and $CoFeBO_4$ have been grown by the flux method. The samples have been characterized by X-ray spectral analysis, X-ray diffraction and X-ray absorption spectroscopy. The X-ray absorption near-edge structure (XANES) and extended X-ray absorption fine structure (EXAFS) spectra have been measured at the $Fe$ and $Co$ $K$-edges over a wide temperature range (6.5 - 300 K). The composition, the charge state and local environment of both Fe and Co atoms have been determined. The effects of Mg substitution by Co on the local structural distortions have been revealed experimentally and the $M-O$ bond anisotropy has been found.

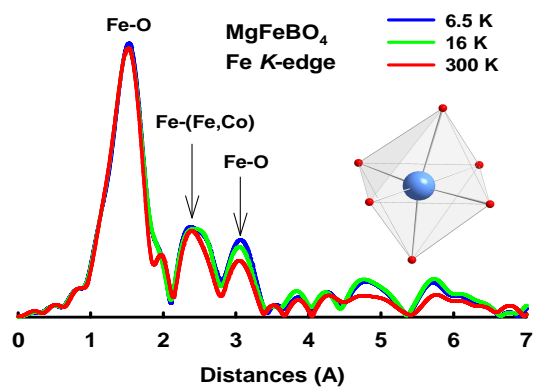

## 1. Introduction

The spectroscopic, magnetic and electronic properties of the oxyborates remain the focus of numerous studies [1-6 and therein]. Special interest is connected with the oxyborates containing the transition and the rare-earth metals. The quasi low-dimensionality, magnetic anisotropy, charge and orbital ordering, spin crossover *etc.* are believed to be the consequences of strong electron correlation and interrelation between spin and orbital degrees of freedom [7, 8].

The warwickites $M^{2+}M^{3+}BO_4$ have been a matter of deep study in the past [9, 10] and recently in the search for strongly correlated systems undergoing charge ordering phase transitions [11, 12]. The crystal structure belongs to either orthorhombic or monoclinic system. The former is attributed to a stretching of an octahedral $M-O-M$ axis, typical of Jahn-Teller (JT) distortion of metal ion. The structure can be described as an assembly of infinite *ribbons* extended along the *c*-axis [13]. The *ribbons* are formed by four *chains* of edge-sharing oxygen octahedra linked in a 2-1-1-2 sequence, where 1 and 2 denote crystallographically nonequivalent metal sites. It is the trigonal borate group $BO_3$ responsible for the bonding between *ribbons* that is the most strongly bonded group of ions in the entire structure.

A variety of hetero-metallic warwickites has been synthesized to date: $MgGaBO_4$[14], $MgFeBO_4$[13], $MgVBO_4$[13], $MgCrBO_4$[13], $CoFeBO_4$[13, 15], $CoCrBO_4$[14], $NiFeBO_4$[16], $Fe_{1.91}V_{0.09}BO_4$[17], $CaInBO_4$[18] are just some representative examples. An important feature of mixed-metal warwickites is the random distribution of metal ions among nonequivalent metal sites giving rise to random short-range AF interactions. It has been found that hetero-metallic warwickites ($M^{2+} \neq M^{3+}$) are usually spin glasses with a relatively low temperature $T_{SG}$ of the magnetic transition [19, 20]. In homo-metallic compounds, the situation is more complicated. Only two homo-metallic warwickites, $Fe_2BO_4$ and $Mn_2BO_4$, have been reliably characterized so far. Both are mixed-valent compounds and demonstrate magnetic, structural and charge ordering transitions [21-24].

Previous investigations on the bimetallic warwickites dealt with structural and magnetic properties of the compounds where different kinds of metal ions were taken in a 1:1 ratio. The crystal structure studies were mainly restricted to room temperature data. The present work is the study of the crystal and local atomic structure around the iron and cobalt atoms in $Mg_{1-x}Co_xBO_4$ warwickite system as a function of temperature (XAS) and $Co$ content (crystallography, XAS). The study includes a detailed structure and electronic state analysis using X-ray single-crystal diffraction. To gain a deeper insight into the valence and coordination state distributions of iron and cobalt ions, X-ray absorption near-edge structure (XANES) and extended X-ray absorption fine structure (EXAFS) experiments at the $Fe$ and $Co K-$edges were conducted. The temperature interval was 6.5 - 300 K. XANES spectra, in particular pre-edge transitions from $1s$ to unoccupied $3d$ states, give the most valuable information on charge distribution. A comparison of intensities and shifts of exact edge positions with respect to those in reference samples provide qualitative information about the coordination and oxidation states of the metal ions [25, 26]. EXAFS spectra provide element-specific local structure parameters [27].

With this respect, the purpose of the present study is to clarify the effects of $Co$ addition on the crystal structure, electronic state and local structure distortions, in conjunction with the variation of the $Fe - O$ and $Co - O$ bond lengths within the $FeO_6$ and $CoO_6$ octahedra.

**2. Experimental procedure**

Single crystals of $Mg - Fe$, $Mg - Co - Fe$, and $Co - Fe$ warwickites were grown by the flux method in the system $Bi_2Mo_3O_{12} - B_2O_3 - CoO - MgO - Fe_2O_3$. The saturation temperature was $T_{sat} \leq 980°C$ and the crystallization interval was $\Delta T_{cr} \geq 30°C$. The flux was heated at 1050°C during 4-6 h and then fast cooled down to $T \approx T_{sat} - (10 - 12)°C$ and subsequently slowly cooled at a rate of (4 - 6)°C/day. The growing process was continued for three days. After that, the product was subjected to etching in 20% aqueous nitric acid. Needle-shaped black crystals with a typical size of $0.5 \times 0.2 \times 5.0$ $mm^3$ were obtained.

Extremely high-purity specimens have been prepared; the composition study was carried out by X-ray spectral analysis using scanning electron - probe microanalyzer (SEPMA) JEOL JXA-8100. The wavelength – dispersive method was used, operating at 15 kV accelerating voltage, 30 nA sample current, 2 – 4 μm beam diameter, and 10 s counting time on peak and 5s on backgrounds. Natural minerals ($Fe_2O_3$, $MgO$, $CoO$, $B_2O_3$) were used as standards. Prior to characterization, samples were polished. The analysis was performed from 20 arbitrarily chosen points across the sample surface. All $Fe$ content was postulated to be $Fe^{3+}$, while $Co$ was assumed to be $Co^{2+}$. The determination of element concentrations was carried out by ZAF quantification procedure. The obtained results show that strict stoichiometry is preserved for all samples. For $Mg - Fe$ warwickites, the ionic ratio $Mg/Fe$ was 0.94/1.06, therefore below we shall use the chemical formula $MgFeBO_4$. For $Co - Fe$ warwickite, the ionic ratio $Co/Fe$ was found to be 0.97/1.03 and the chemical formula will be $CoFeBO_4$. In the $Mg - Co - Fe$ warwickite, the content of $Mg$ and $Co$ was found to be approximately equal and half of the $Fe$ content, so we formulate this sample as $Mg_{0.5}Co_{0.5}FeBO_4$.

X-ray crystallographic study was carried out with a SMART APEX II diffractometer with graphite-monochromatic MoKα radiation. The SHELXL – 97[28] software package was used to solve the structure and refine it with the full-matrix least-$k$ squares method on $F^2$.

The XAS measurements were performed at the Structural Materials Science beamline of the National Research Center "Kurchatov Institute" (Moscow) in the transmission mode. XAS measurements were carried out at temperatures ranging from 6.5 to 300 K using a SHI closed-cycle helium refrigerator (Japan). The storage ring operated at an electron energy of 2.5 GeV and an average electron current of about 80 mA. For the selection of the primary beam photon energy, a Si (111) channel-cut monochromator was employed, which provided an energy resolution $\Delta E/E \sim 2 \cdot 10^{-4}$. Primary and transmitted intensities were recorded using two independent ionization chambers filled with appropriate $N_2/Ar$ mixtures. The energy was calibrated using corresponding metal foils.

The EXAFS spectra were collected at the $Fe$ and $Co K$ absorption edges using optimized scan parameters of the beamline software. The $\Delta E$ scanning step in the XANES region was about 0.5 eV, and scanning in the EXAFS region was carried out at a constant step on the photoelectron wave number scale with $\Delta k = 0.05 \text{ Å}^{-1}$, giving energy step of order of 1.5 eV. Single-crystalline samples for XAS measurements were ground to fine powders and then spread uniformly onto a thin adhesive kapton film and folded into several layers to give an absorption edge jump around unity.

The EXAFS spectra $\mu(E)$ were normalized to an edge jump and the absorption coefficient of the isolated atoms $\mu_0(E)$ was extracted by fitting a cubic-spline-function to the data. After subtraction of the atomic background, the conversion from $E$ to $k$ scale was performed. Structural parameters for the starting model were taken from the X-ray single-crystal diffraction analysis. The $k^3$-weighted EXAFS function $\chi(k)$ was calculated in the intervals $k$= 2 - 15.3 Å$^{-1}$ ($CoFeBO_4$) and $k = 2 – 12.2$ Å$^{-1}$($MgFeBO_4$ and $Mg_{0.5}Co_{0.5}FeBO_4$) using a Kaiser-Bessel window (local order peaks were clearly distinguishable against background up to 7 Å). The EXAFS structural analysis was performed using theoretical phases and amplitudes as calculated by the FEFF-8 package [29], and fits to the experimental data were carried out in the $R$-space with the IFFEFIT package [30]. EXAFS analysis of the warwickite systems remains complicated due to the presence of two nonequivalent metal sites in the crystal structure giving rise to strongly overlapped contributions of coordination shells. Therefore the EXAFS analysis is restricted to the first oxygen shell $M - O$.

### 3. Results and discussion
### 3.1. X-ray diffraction

The crystallographic data for $Mg - Fe$, $Mg - Co - Fe$, and $Co - Fe$ warwickites are summarized in Table 1 and in Tables SI1 and SI2 of Supporting Information, Ref. [31]. The compounds crystallize in an orthorhombic system. The lattice parameters vary linearly as the cobalt content increases (thus obeying the Vegard's law). The $c$ - parameter increases while $a$ and $b$ decrease. All parameters are in a good agreement with those reported earlier[13]. The metal ions occupy two nonequivalent crystallographic sites both belonging to general $4c$ Wyckoff positions. These positions are indicated as M1 and M2 (see Fig. SI1). The boron ions reside in only one position and $O$ ions are distributed among four distinct positions.

The general features of the crystal structure are typical for warwickites[32]: the metal ions are surrounded by oxygen octahedra. These octahedra are linked by sharing their edges thus forming four - octahedra flat ribbons extending along the $c$- axis (Fig. 1). The row consisting of four octahedra stacked in the sequence 2 – 1 – 1 - 2 is located across the ribbon. The coordination octahedra around the M2 position form the outer columns of the ribbon and the octahedra around the M1 position form the inner two columns. The planar trigonal borate group ($BO_3$) located in the voids between the ribbons are attached to them by corner sharing. The mean $B - O$ bond length and the mean $O - B - O$ bond angle are in a good agreement with a trigonal planar geometry in all samples.

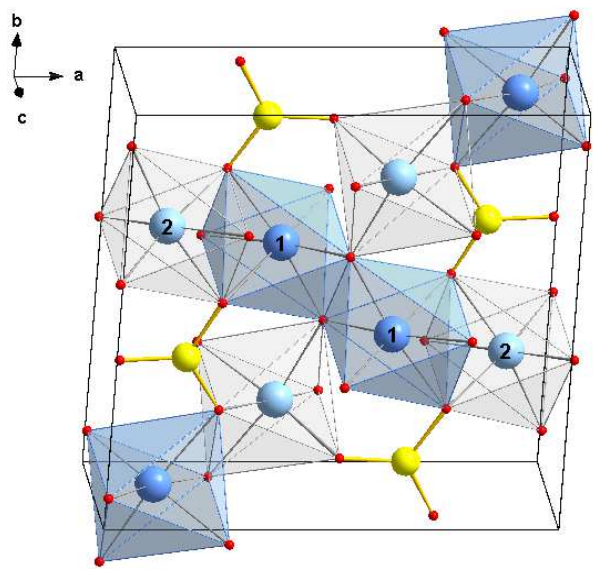

**Fig.1** The schematic structure of warwickite. The metal cations have octahedral coordination, where the edge-sharing octahedra form ribbons. Coordination octahedra around the M1 position (labeled 1) are dark and those around the M2 position (labeled 2) are light. The boron atom positions drawn as yellow circles have trigonal coordination. The unit cell and axes directions are shown.

**Table 1** Crystal data and refinement results for $Mg-Fe$, $Mg-Co-Fe$ and $Co-Fe$ warwickites.

|  | $Mg-Fe$ | $Mg-Co-Fe$ | $Co-Fe$ |
|---|---|---|---|
| Formula weight (g mol$^{-1}$) | 154.97 | 172.27 | 189.59 |
| Crystalsystem | | orthorhombic | |
| Space-group | | $Pnma$ (62) | |
| Unit cell parameters (Å) | | | |
| $a$ | 9.2795(10) | 9.2449 | 9.2144 |
| $b$ | 9.4225(10) | 9.3898 | 9.3651 |
| $c$ | 3.1146(3) | 3.1185 | 3.1202 |
| Unit cell volume (Å$^3$) | 272.33(5) | 270.71(6) | 269.25 |
| Z | | 4 | |
| Calculated density (gcm$^3$) | 3.78 | 4.22 | 4.68 |
| Radiation | | MoKα | |
| Wavelenght λ (Å) | | 0.71073 | |
| Temperature ($K$) | | 296 | |
| Crystal shape | | Needle (along $c$) | |
| Abs. coefficient ($mm^{-1}$) | 2.785 | 5.764 | 5.672 |
| F(000) | 150 | 204 | 180 |
| $\Theta$ range ($deg$) | 3.08 - 29.44 | 3.09 – 29.74 | 3.10 - 29.75 |
| Limiting indices | | $-12 \leq h \leq 12$ $-4 \leq k \leq 4$ $-12 \leq l \leq 12$ | |
| Extinctioncoefficient | 0.123(6) | 0.125(7) | 0.256(14) |
| Data / restraints / parameters | 437 / 0 / 44 | 439 / 0 / 46 | 437 / 0 / 44 |
| Goodness-of-fit on $F^2$ | 1.187 | 1.133 | 1.223 |
| Final R indices | | | |
| R1 | 0.0187 | 0.0228 | 0.0290 |
| wR2 | 0.0459 | 0.0575 | 0.0759 |

The ionic distributions can be studied by the calculation of bond valence sums ($BVS$) at a specific site $BVS_i = \sum_j exp(r_0 - r_{ij})/B$, with $r_{ij}$ being the bond length between two atoms, $r_0$ and $B$ are empirical parameters tabulated in Ref's 33, 34. Using the parameters for $Mg^{2+}$, $Co^{2+}$, $Co^{3+}$, $Fe^{2+}$, and $Fe^{3+}$ the bond valence sums ($BVS$) for the M1 and M2 positions are listed in Table 2. With the $Mg^{2+}$ and $Co^{2+}$ parameters the corresponding $BVS$ values lie in the range of 2.14 - 2.23 and 2.18 – 2.22 valence units (v.u.), respectively. Usually, bond valence sums contain variations of about 10 % even in typical compounds, such as $CoO$ and $Co_2O_3$, which can be attributed to inaccuracy of the interatomic distances. However, the BVS values for $Co^{3+}$ (~1.89) and $Fe^{2+}$ (~2.46) are by far different from the expected values 3 and 2, respectively. Therefore, it can be considered that in the warwickites under investigation, the metal ions correspond to the following charge states: $Co$ and $Mg$ are divalent and $Fe$ is trivalent. The small reduction of the BVS value at the M2 position in comparison with the M1 one in the $Mg - Fe$ and the $Mg - Co - Fe$ samples can be an indication that $Mg$ has a preference for this position (M2).

Table 2 The estimated empirical bond valence sums for the metal ions (v.u.).

|  | $Mg - Fe$ | | $Mg - Co - Fe$ | | $Co - Fe$ | |
|---|---|---|---|---|---|---|
|  | M1 | M2 | M1 | M2 | M1 | M2 |
| $Mg^{2+}$ | 2.23 | 2.14 | 2.21 | 2.19 | | |
| $Co^{2+}$ | | | 2.21 | 2.18 | 2.19 | 2.22 |
| $Co^{3+}$ | | | 1.90 | 1.88 | 1.89 | 1.91 |
| $Fe^{2+}$ | 2.49 | 2.39 | 2.47 | 2.44 | 2.46 | 2.49 |
| $Fe^{3+}$ | 2.66 | 2.56 | 2.65 | 2.61 | 2.63 | 2.66 |
| $B^{3+}$ | 2.89 | | 2.93 | | 2.96 | |

The sum of the metal ion occupancies at each of two distinct crystallographic positions was fixed to 1. The refinements of the occupancies of $Fe$ and $Mg$ at the M1 and M2 positions in the $Mg - Fe$ warwickite indicated 72% ($Fe$) + 28% ($Mg$) and 28% ($Fe$) + 72% ($Mg$), respectively. Thus, the M1 position can be treated as preferably occupied by $Fe$ ions. The average valences at M1 and M2 positions for $Mg - Fe$ warwickite are found to be 2.54 and 2.26, respectively. This is roughly consistent with the valence distribution in a warwickite (for example, in $Mg - Sc$ warwickite the charges are +2.76 and +2.24 for M1 and M2 position, respectively) [35]. The small difference in the scattering power of $Fe$ and $Co$ does not allow unambiguous determination of the occupancies of M1 and M2 positions in the $Mg-Co-Fe$ and $Co-Fe$ samples.

Table SI2 gives some important interatomic distances. Note that the octahedron $M1O_6$ is smaller than $M2O_6$ in the $Mg-Fe$ and the $Mg-Co-Fe$ warwickites, as deduced from the average $\langle M-O \rangle$ distances. As the $Co$ content increases, the average $\langle M-O \rangle$ bond distances vary. The $\langle M1-O \rangle$ bond distance increases, while the $\langle M2-O \rangle$ bond distance decreases. As a consequence, the $M1O_6$ octahedron becomes larger than $M2O_6$ ones in the $Co-Fe$ warwickite. The shorter $\langle M-O \rangle$ distances provide stronger 3$d$ repulsion and lead to apparently increased oxidation state. It means that trivalent Fe ions would prefer to occupy the smaller octahedra $M1O_6$ in the $Mg-Fe$ and $Mg-Co-Fe$ warwickites, while they would reside at the $M2O_6$ octahedra in the $Co-Fe$ warwickite.

The $M1O_6$ and $M2O_6$ octahedra undergo high-symmetry distortions such as a breathing mode and the low-symmetry rhombic and trigonal distortions. Normal coordinates $Q_\alpha$ ($\alpha = 1, 2, …3N – 3$; $N$ – number of ligands) are linear combinations of the Cartesian coordinates of oxygen, and classified according to the irreducible representations of the coordination complex symmetry (Tables 3 and SI3), in terms of the $O_h$ symmetry group. The $Q_1$ coordinate describes high-symmetry distortions. The other normal coordinates correspond to low – symmetry distortions like rhombic (or JT, $Q_2$ and $Q_3$) and trigonal ($Q_4, Q_5, Q_6$) ones. The $Q_3$ coordinate presents tetragonal octahedral distortion along the $z$– axis, whereas the $Q_2$ ones corresponds to the distortions with rhombic symmetry. The metal – oxygen bond lengths are distributed so that four bonds (O3, O4 for M1 site and O2, O1 for M2 one) are roughly coplanar and the two other ones (O2, O4 for M1 and O3, O4 for M2) are axial (see Fig. SI1). The JT-like compression can be seen to involve a O2-M1-O4 and O3-M2-O4 axes. There is a stretching of the other two axes, with the largest distances among two M-O bonds (~2.165 and 2.156 Å for $x = 0.0$ and 1.0, respectively). The compression is significant for M1 site in $Mg-Fe$ and for M2 site in $Co-Fe$, as seen from the planar/axial average radii ratio (2.079/2.045 Å for $Mg-Fe$) and (2.085/2.035 Å for $Co-Fe$). In $MgFeBO_4$ the rhombic distortion of the $M2O_6$ octahedra is larger than that of the $M1O_6$, while the tetragonal distortion is more pronounced for the $M1O_6$ octahedron. The $Co$ addition gives rise to a rapid decrease in the axial average radius of the $M2O_6$ octahedra (from 2.074 Å for $x = 0$ to 2.035 Å for $x = 1.0$) as compared to one of the $M1O_6$ ones (from 2.045 Å for $x = 0$ to 2.041 Å for $x= 1.0$). That leads to a situation when both types of the low-symmetry distortions $(Q_2, Q_3)$ are prevailing for M2 site for the $Co - Fe$ warwickite. The rhombic distortion of the $M1O_6$ octahedron increases by approximately one order of magnitude as $Co$ content increases, as can be seen from comparison of the $Q_2$ coordinates (Table 3). The tetragonal distortion shows a significant growth for the $M2O_6$ octahedra with an increase in the $Co$ concentration. Thus, the JT-like distortion is expected to increase as the cobalt content increases. Besides, the trigonal distortions are comparable for both oxygen octahedra and are weakly affected by the $Co$-substitution.

**Table 3** The ligand's displacement $(\text{Å})$ for metal ions in the $M1O_6$ and $M2O_6$ octahedral complexes. The $R_0$ is the $M - O$ distance in the undistorted octahedron, that are accepted such in order to provide a zero value of $Q_1$.

| Normal coordinates | $Mg - Fe$ | | $Mg - Co - Fe$ | | $Co - Fe$ | |
|---|---|---|---|---|---|---|
| | M1 | M2 | M1 | M2 | M1 | M2 |
| $R_0$ | 2.0487 | 2.0701 | 2.05015 | 2.0627 | 2.05235 | 2.05681 |
| $Q_2$ | 0.002 | 0.0133 | 0.0201 | 0.0143 | 0.0193 | 0.0149 |
| $Q_3$ | -0.0131 | -0.0088 | -0.0169 | -0.0319 | -0.0249 | -0.0496 |
| $Q_4$ | 0.2414 | 0.1761 | 0.2405 | 0.1627 | 0.2388 | 0.1447 |
| $Q_5$ | 0.2407 | 0.1725 | 0.2398 | 0.1594 | 0.2384 | 0.1416 |
| $Q_6$ | -0.2893 | -0.2534 | -0.2896 | -0.2597 | -0.2833 | -0.2633 |
| $Q_7$ | 0.0601 | -0.0829 | 0.053 | -0.0765 | 0.0508 | -0.0718 |
| $Q_8$ | 0.069 | -0.0921 | 0.0608 | -0.0854 | 0.0582 | -0.0805 |
| $Q_9$ | -0.026 | -0.0576 | -0.0163 | -0.0645 | -0.0132 | -0.0710 |
| $Q_{10}$ | 0.0479 | -0.024 | 0.0421 | -0.0244 | 0.0384 | -0.0238 |
| $Q_{11}$ | 0.055 | -0.0267 | 0.0484 | -0.0272 | 0.0441 | -0.0266 |
| $Q_{12}$ | 0.0459 | -0.034 | 0.0421 | -0.0327 | 0.0451 | -0.0319 |
| $Q_{13}$ | 0.0077 | 0.0489 | 0.0066 | 0.042 | 0.0036 | 0.0382 |
| $Q_{13}$ | -0.0089 | -0.0543 | -0.0076 | -0.0469 | -0.0041 | -0.0428 |
| $Q_{15}$ | 0 | 0 | 0 | 0 | 0 | 0 |

We have estimated the octahedral deformation by means of the electric field gradient (EFG) calculation. The EFG generated by the oxygen octahedron on the metal sites M1 an M2 can be expressed as following: $G_{\alpha\beta} = 2e \sum_i \frac{3cos^2\varphi_i - 1}{r_i^3}$, $\varphi_i$ is the angle between the principal octahedra axis and the direction towards the $i$-th oxygen anion, $r_i$ is the distance between the cation and the $i$-th oxygen anion. The main component $(V_{zz})$ of the $G_{\alpha\beta}$- tensor for both metal sites are listed in Table 4. The EFG created at M1 in the $Mg - Fe$ warwickite is considerably higher than the one at the M2 site showing that site M1 is ~2.1 times more distorted than site M2. As the $Co$-content increases, the $V_{zz}$ shows a gradual growth for the inner $M1O_6$ and a rapid increase for the outer $M2O_6$ oxygen octahedra. So, from the EFG calculation, we conclude that $Co$ substitution induces an increase in degree of distortion of the coordination octahedra around the metal ions. The EFG principal axis lies along the M1 – O2 and M2 – O3 bonds for the $M1O_6$ and $M2O_6$ octahedra, respectively. The directions of the EFG principal axes are alternated from site to site in the row 2 – 1 – 1 – 2, which indicates that there is inversion of the principal axis for every second ion along the ribbon substructure (Fig. 2).

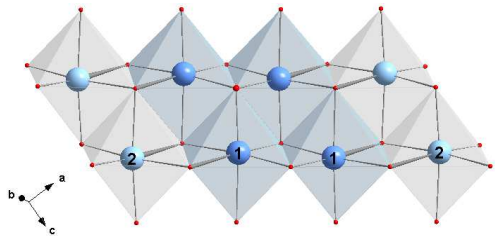

**Fig. 2** Two 2-1-1-2 rows of the infinite ribbon are shown. The direction of the principal axes of the octahedrons $M1O_6$ and $M2O_6$ are highlighted in bold.

**Table 4** The main component $V_{zz}$ of EFG tensor for $M1O_6$ and $M2O_6$ oxygen octahedra $(e \cdot Å^3)$.

|  | M1 | M2 |
|---|---|---|
| $MgFeBO_4$ | 0.0928 | 0.0438 |
| $Mg_{0.5}Co_{0.5}FeBO_4$ | 0.0954 | 0.0696 |
| $CoFeBO_4$ | 0.1040 | 0.0873 |

The main results of the structural study on $Mg-Fe$, $Mg-Co-Fe$ and $Co-Fe$ warwickites may be summarized as follows: 1) $Co$ and $Mg$ enter into the warwickite structure in the divalent state, and $Fe$ in the trivalent state; 2) both (M1 and M2) positions are occupied by a mixture of $Mg$, $Co$, and $Fe$ atoms; 3) the trivalent $Fe$ ions prefer smaller octahedra: $M1O_6$ in the $Mg-Fe$ and $Mg-Co-Fe$ warwickite, and the $M2O_6$ one in the $Co-Fe$ compound; 4) both octahedra are compressed along one of the nominal 4-fold axis and the $Co$ addition increases the octahedral distortion; 5) the principal axis of the octahedra is inverted for every second metal ions along the ribbon structure.

### 3.2. X-ray absorption spectroscopy
#### 3.2.1. XANES spectra

Figure 3 shows normalized $FeK$-edge XANES spectra and their first derivatives of the warwickites under study recorded over a wide temperature range. The $FeO$ and $Fe_2O_3$ have been used as references for the $Fe^{2+}$ and $Fe^{3+}$ charge states. There is a strong similarity in the spectral line shapes at different temperatures. A weak pre-edge absorption feature ($A$) at ~ 7115 eV, followed by a weak shoulder on a rising absorption curve (the absorption edge is at ~7124 eV) which culminates in a strong peak in the vicinity of ~7132 eV ($B$). This strong peak has been assigned to a dipole-allowed transition $1s-4p$ and the pre-edge feature just below the threshold as a dipole-forbidden transition $1s - 3d$, which has a non-zero probability due to a partial $p - d$ mixing and quadrupole contribution. The difference in the energy position of the main absorption edge for the samples and $FeO$ reference can be seen. The main edge position for warwickites is found to be ~3 eV higher than that for $Fe^{2+}$ compounds, as deduced from the maximum of the derivative spectra in Fig. 3(b), (d), and (f), but very close to that of $Fe^{3+}$ compound thus indicating that the valence state of the $Fe$ is mainly +3. The pre-edge $1s - 3d$ transition probability is related to the coordination symmetry and to the occupancy of the $3d$ shell of the transition metal. As the iron ions were found to be in the trivalent state, the transition is closely associated with the presence of the local inversion symmetry of the first coordination shell. In an ideal octahedral symmetry, the $p - d$ mixing is symmetry-forbidden. For non-centrosymmetric environments around the $Fe$ ion, *i.e.*, in distorted $FeO_6$ octahedra, this transition gains some intensity. The weak intensity of the pre-edge feature is in agreement with the 6-coordinated $Fe$ being located in relatively weakly distorted sites.

The room temperature $FeK$-edge spectra of all three samples are plotted in Fig. 4. The curve's behavior is common for all orthorhombic warwickites. As shown in the insets of Fig. 4, the intensity of the peak $A$ experiences no changes while the intensity of the peak $B$ slightly reduces with increasing $Co$ content. This effect can be explained in terms of the difference in the degree of the orbital mixing which arises from the variation of local structure upon $Co$ doping. The main-edge position is independent of both Co content and temperature. Thus, the XANES results at the $FeK$-edge are consistent with the previous finding that $Fe$ ions enter the warwickite structure in the trivalent state predominantly and are located inside the slightly distorted octahedral sites.

The normalized $CoK$-edges XANES spectra of the $Co$-containing warwickites measured over a wide temperature range are shown in Fig. 5. Also the room temperature spectra of $CoO$ and $Co_2O_3$ are included as references for the $Co^{2+}$ and $Co^{3+}$ charge states respectively. The warwickites spectra are characterized by a low-intensity peak at 7709 eV, a shoulder at 7715 eV, and a steeply rising edge (the main absorption edge is at 7719 eV), that leads to a series of well-defined peaks at 7709 (*A*), 7725 (*B*) and 7738 eV (*C*), and to others at higher energy in the EXAFS region. The inflection point associated with each of these features shows up more clearly in the derivative spectrum (bottom panels of Fig. 5). The weak intensity pre-edge peak at ~ 7709 eV correspond to the $1s$ - $3d$ absorption process. The main peak (7725 eV) corresponds to the $1s - 4p$ transition. The main edge energy (7719 eV) is very close to that $CoO$ (7720 eV), but prominently different from that of $Co_2O_3$ (7727 eV), thus indicating the prevailing divalent state of cobalt. Since there is no shift of the main edge absorption as $Co$ content increases (Fig. 6), we conclude that the valence of cobalt is independent of cobalt content. Both samples exhibit same intensity in the pre-edge feature at ~ 7709 eV, and in the peak $B$ at 7225 eV. It reflects the fact that $Co$ environment remains six-coordinate and distorted as $Co$ content increases. This conclusion is supported by the crystal structure refinement where tetragonal distortion of both octahedra was clearly revealed.

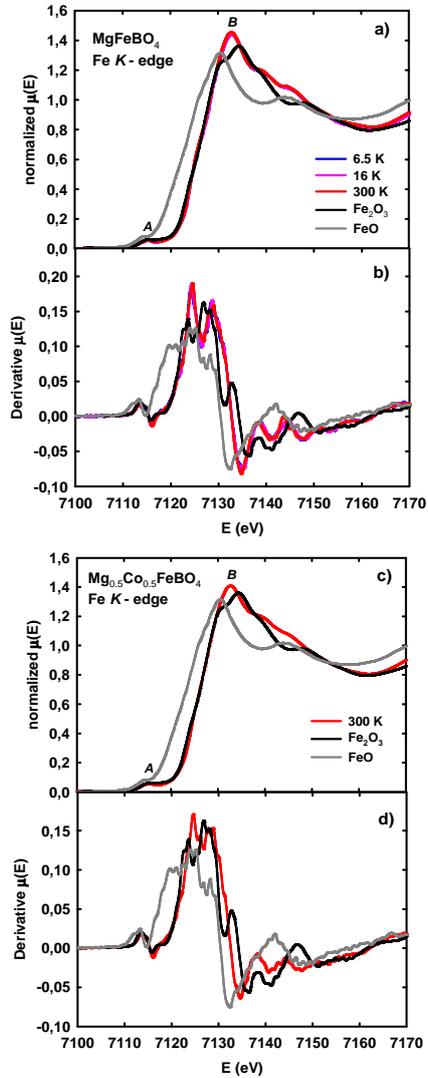

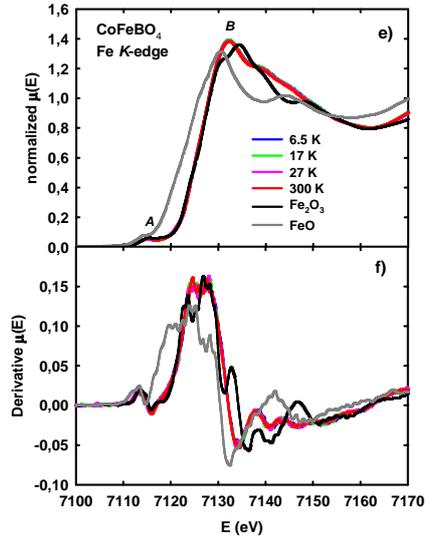
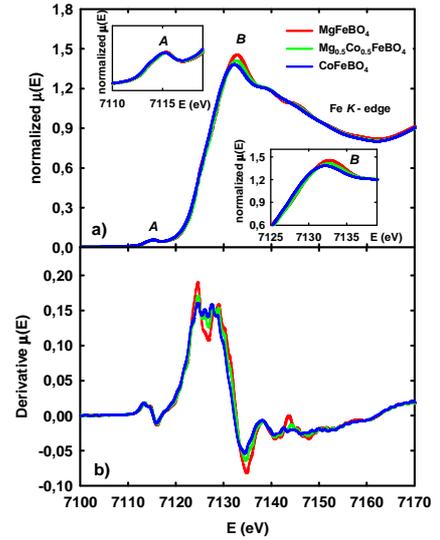

**Fig. 3** Normalized $FeK$-edge spectra of the warwickites at different temperatures and the reference samples $FeO$ and $Fe_2O_3$ at room temperature: a) $MgFeBO_4$, c) $Mg_{0.5}Co_{0.5}FeBO_4$, e) $CoFeBO_4$. b), d), f) First derivatives of these spectra (same colours).

**Fig. 4** a) Normalized Fe $K$-edge spectra of $MgFeBO_4$, $Mg_{0.5}Co_{0.5}FeBO_4$ and $CoFeBO_4$ at room temperature. Top inset: $Co$-content independent pre-edge peak. Bottom inset: the $B$ peak intensity dependence on the $Co$-content. b) First derivatives of the spectra.

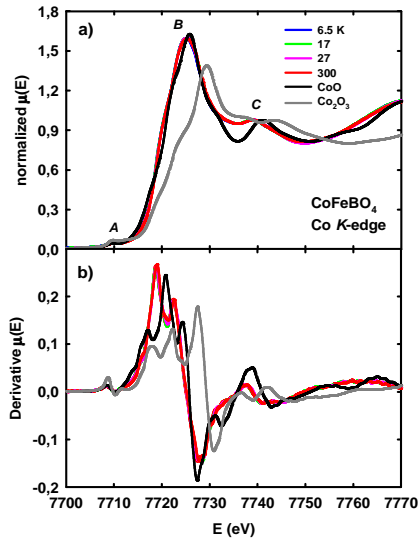
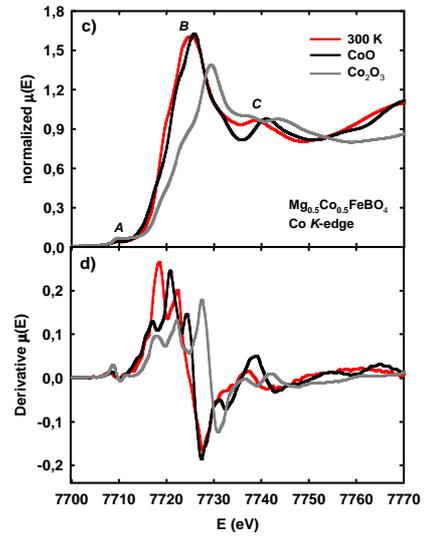

**Fig. 5** Normalized $CoK$-edge spectra of warwickites and the reference samples $CoO$ and $Co_2O_3$ at room temperature: a) $CoFeBO_4$, c) $Mg_{0.5}Co_{0.5}FeBO_4$. b), d) First derivatives of these spectra (same colours).

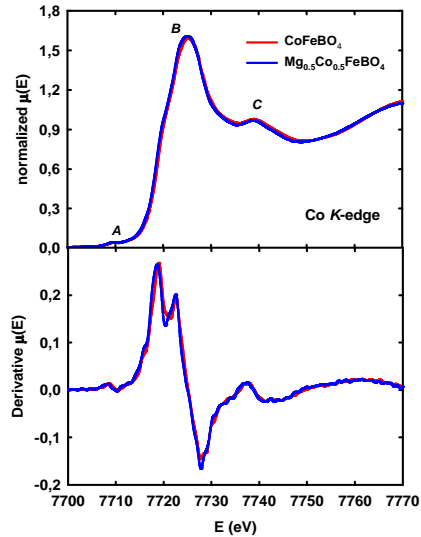

**Fig. 6** Normalized $CoK$-edge spectra of $Mg_{0.5}Co_{0.5}FeBO_4$ and $CoFeBO_4$ at room temperature showing no dependence of XANES spectra on the $Co$ content.

### 3.2.2. EXAFS spectra

According to the X-ray crystallographic data, the average $MO_6$ octahedron becomes smaller and progressively distorted with $Co$ doping. We have examined $Co$ substitution effects on the local structural distortion by analyzing the $Co$ and $FeK$-edge EXAFS spectra.

EXAFS spectra of three orthorhombic warwickites were measured at variable temperatures from 6.5 K up to 300 K. Figures 7 and 8 show the Fourier transforms (FT) of the EXAFS functions for $MgFeBO_4$, $Mg_{0.5}Co_{0.5}FeBO_4$ and $CoFeBO_4$ at the $Fe$ and $CoK$-edges, respectively. Three main features can be observed. The first one is the peak at about 1.5 Å corresponding to the first oxygen coordination shell $Fe-O$ (Fig. 7). This single peak is split into two distinct peaks (a doublet) at ~1.28 and 1.61 Å corresponding to the first-shell $Co-O$ bonds at the $CoK$- edge spectra (Fig. 8). Second feature between 2 and 3 Å is mainly related to the nearest metal neighbors $(Fe, Co)$ along the $M-O-M$ chains and the third one lies between 3 and 4 Å and includes contributions from the second-shell oxygen atoms $(M-O)$ mainly. Peaks observed at even longer distances correspond to contributions from higher-shell neighbors. The temperature evolution of the $FeK$-edge EXAFS spectra for the two end warwickites is presented in Fig. 7(a) and (c). Both samples show a decrease in the first peak height as temperature increases. A similar trend is observed for the second peak. The intensity of the first coordination shell peak decreases more progressively for $CoFeBO_4$ in comparison with that for $MgFeBO_4$. The second peak for $CoFeBO_4$ becomes less asymmetric as temperature increases. The weak temperature dependence of the intensity of the first oxygen shell found for $MgFeBO_4$ can indicate a nearly constant distortion of the $FeO_6$ octahedra within the temperature range studied. Moreover, the intensity decrease of the first oxygen shell peak is more pronounced as $Co$ content increases. It can be seen from the comparison of $MgFeBO_4$ and $CoFeBO_4$ FT's at room temperature (Fig. 9(a)). The intensity of the first-shell is smaller for the $CoFeBO_4$. This fact indicates a more distorted local structure around the $Fe$ atom for this compound.

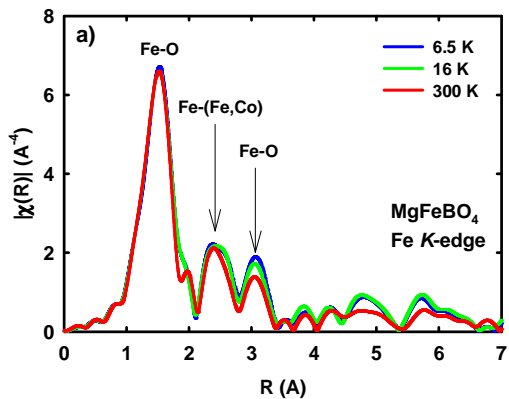

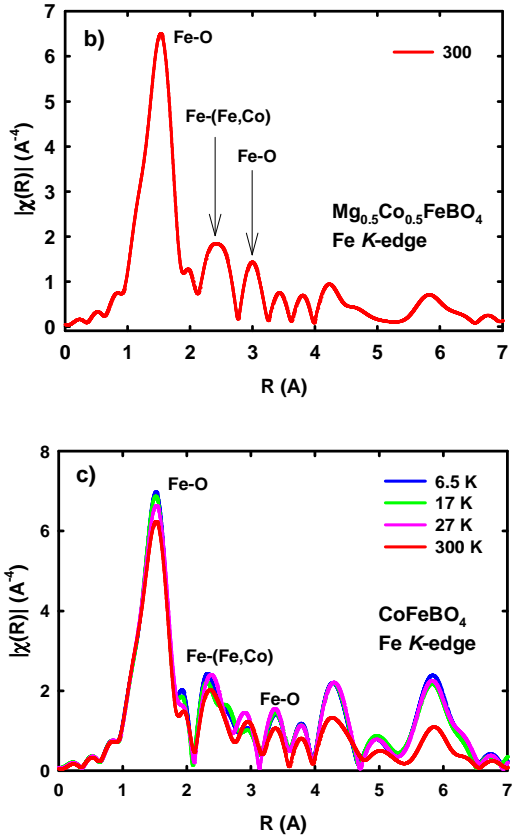
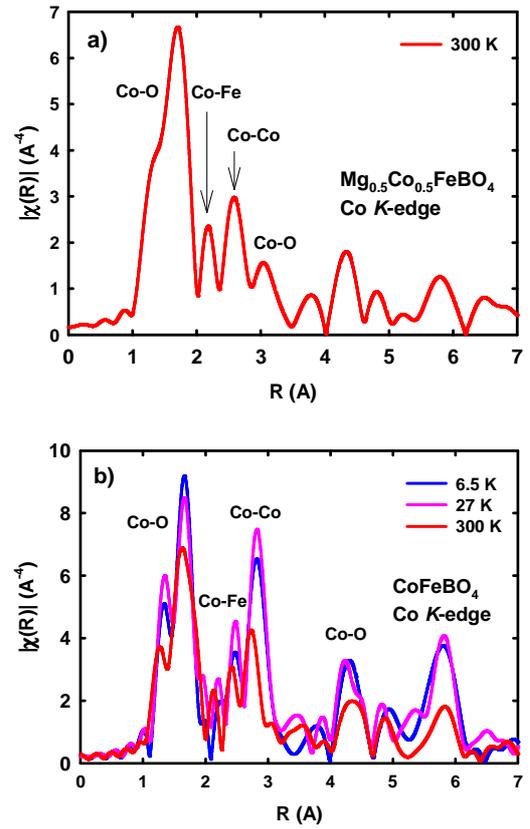

**Fig. 7** FT modulus of the $k^3$-weighted EXAFS spectra of the warwickites at the $FeK$-edge as a function of temperature: a) $MgFeBO_4$; b) $Mg_{0.5}Co_{0.5}FeBO_4$; c) $CoFeBO_4$.

**Fig. 8** FT modulus of the $k^3$-weighted EXAFS spectra of the warwickites at the $CoK$-edge as a function of temperature: a) $Mg_{0.5}Co_{0.5}FeBO_4$; b) $CoFeBO_4$.

The temperature dependence of the EXAFS spectra at the $CoK$−edge has shown that the intensity of the doublet corresponding to the first oxygen shell $Co - O$ for $CoFeBO_4$ significantly decreases with increasing tempera ture, indicating the distortion of the coordinated octahedra around the $Co$ ion (Fig. 8(b)). The second peak demonstrates a similar temperature dependence. The comparison of room temperature FT modulus of $CoK$-edge EXAFS spectra for $Mg_{0.5}Co_{0.5}FeBO_4$ and $CoFeBO_4$ is presented in Fig. 9(b). The radial distribution function demonstrates that the asymmetric single peak with a wide shoulder corresponding to the first-shell $Co - O$ bonds of $Mg_{0.5}Co_{0.5}FeBO_4$ is gradually split into two distinct peaks (a doublet) with increasing $Co$-doping content. This indicates a gradual increase in the local structural distortion of the $CoO_6$ octahedron with the $Co$ substitution.

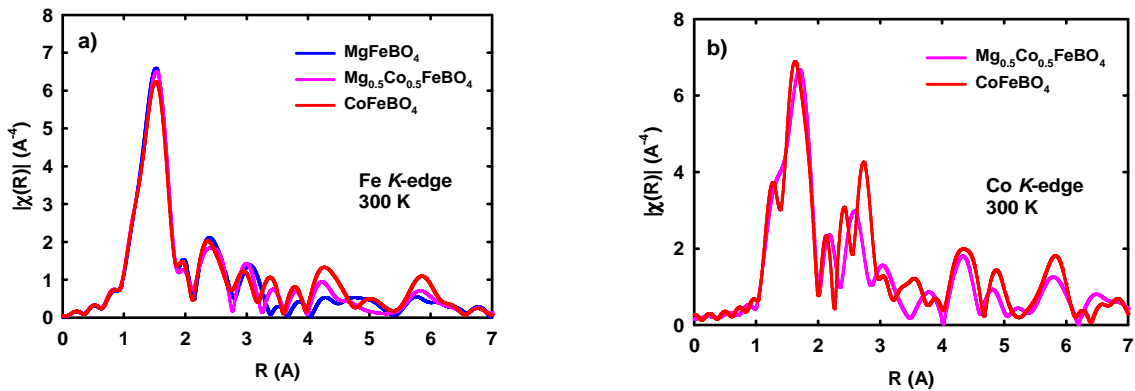

**Fig. 9** The room temperature FT modulus of the $k^3$-weighted EXAFS spectra of the $MgFeBO_4$, $Mg_{0.5}Co_{0.5}FeBO_4$, and $CoFeBO_4$: a) $FeK$-edge, b) $CoK$-edge.

The main results of the quantitative structural analysis for the first coordination shell of the warwickites are summarized in Table SI4 of Supplementary Materials [31]. The amplitude reduction factor $S_0^2$ was fixed to 0.85. The variables in the fits were the interatomic distances $R_{M-O}$ and Debye-Waller (DW) factor $\sigma^2$. For each sample fits at different temperatures were performed with a fixed value of the threshold energy $E_0$ = 2 eV obtained from the fit at the lowest temperature. The comparison between best fit and experimental spectra in term of the FT modulus of the $k^3$-weighted EXAFS signals and the real part of the Fourier-filtered spectra in $k$-space for CoFeBO$_4$ are shown in Fig. SI2. Similar agreements were found for the other warwickites. We have adopted the orthorhombic crystallographic structure at high temperature to calculate theoretical amplitudes and phases for each scattering path up to 7.0 Å, including the first Fe,Co-O coordination shell. We have been able to distinguish four short $R^{short}$ and two long $R^{long}(Fe, Co - O)$ bonds. The temperature dependence of the average $Fe - O$, $Co - O$ inter-atomic distances and its Debye–Waller (DW) factors are compared in Fig. 10(a) and (b) respectively. Both $Fe - O$ and $Co - O$ average inter-atomic distances show no dependence on the Co content. The average $Fe - O$ distance in both $MgFeBO_4$ and $CoFeBO_4$ remains almost constant in the whole temperature range within the experimental error. For $MgFeBO_4$, the DW factor of the average $Fe - O$ distance does not show strong changes with decreasing temperature. These results indicate that the local distortion of the $FeO_6$ octahedra in $MgFeBO_4$ is almost temperature-independent in the whole temperature range. On the contrary, a rather pronounced temperature dependence of the DW factor of the average $Fe - O$ distances was found for $CoFeBO_4$. The DW factor of each scattering path includes contributions from both thermal vibrations and local distortions. The thermal vibration contribution decreases upon cooling down. A large DW factor obtained for the high-temperature phase indicates that the local environment around the $Fe$ atom is locally distorted, with the dynamical contribution prevailing. The DW factor of the $Co - O$ first shell distances shows weak variation with temperature. The value of $\sigma^2$ factor increases for both average distances Fe-O and Co-O with Co content. Thus, the Co addition induces an increase in the local structure distortion of both $FeO_6$ and $CoO_6$ octahedra.

The EXAFS spectra are very useful to distinguish between individual contributions of each type of surrounding atoms that otherwise are overlapped in the X-ray crystallographic data. The warwickite structure has $Mg$, $Co$ and $Fe$ atoms distributed over two nonequivalent local environments. According to Shannon [37], average $Fe - O$ bond lengths 2.14 and 2.02 Å, respectively, are expected for octahedrally coordinated $Fe^{2+}$ and $Fe^{3+}$ ions. The average $Fe - O$ bonds distance of the $FeO_6$ octahedra deduced from EXAFS data is ~2.03 Å, strongly implying purely trivalent state of the iron ions. A larger value of the average $Co - O$ distance as compared to the $Fe - O$ one is obtained for both $Mg_{0.5}Co_{0.5}FeBO_4$ and $CoFeBO_4$. This increase is correlated with the decrease in the formal valence state of the octahedral $Co$ ion, namely, $Co^{2+}$ instead $Fe^{3+}$. The average $Co-O$ bond lengths for both $Co$-warwickites (~2.10 Å) are close to 2.09 Å expected for the octahedrally coordinated $Co^{2+}$ ion [37]. So, from the EXAFS study we conclude that iron and cobalt enter into the warwickite structure in the trivalent and divalent states, respectively. The $Co^{2+}$ atoms with larger ionic radii ($r_i = 0.72$ Å) replace the $Fe^{3+}$ ones ($r_i = 0.645$ Å) at the octahedral M1 site thus increasing the average M1-O bond lengths. This increase is correlated with the shortening of average $M2 - O$ bonds suggesting that $Fe^{3+}$ ions move to the M2 octahedral site. The fact that $Co$ atoms could replace $Fe$ ones at the octahedral site in $CoFeBO_4$ would induce an apparent increase of the DW factor.

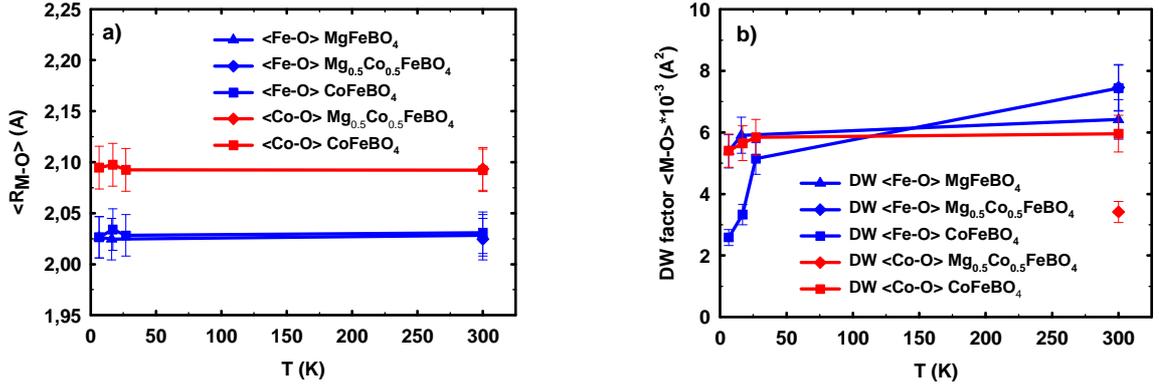

**Fig. 10** Temperature dependences of a) average inter-atomic $Fe - O$ and $Co - O$ distances and b) DW factors for the average $Fe - O$ and $Co - O$ distances.

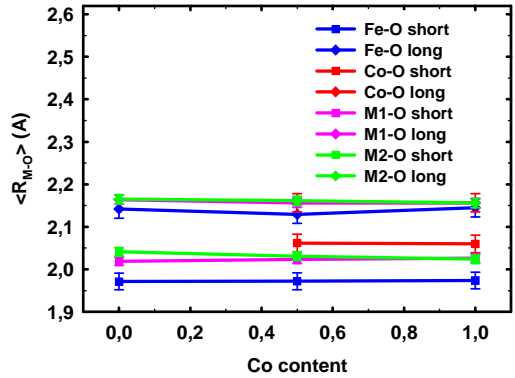

**Fig. 11** The variation of the two distinct average bond distances $M - O$ with $Co$ content obtained from X-ray crystallographic and EXAFS analysis. $T = 300$ K.

The variation of the two distinct distances as a function of $Co$-doping content obtained from the structural refinement and from the EXAFS study is presented in Figure 11. Two oxygen ions are located at a distances of $R_{Fe-O}^{long} = R_{Co-O}^{long} = 2.14 – 2.16$ Å forming "long bonds" that approach the value of the long bond $R_{M-O}^{long} = 2.15 - 2.16$ Å given for the orthorhombic structure from diffraction studies. The four remaining oxygen ions are at the average distances of $R_{Fe-O}^{short} = 1.97$ Å and $R_{Co-O}^{short} = 2.06$ Å belonging to the "short bonds". These distances are close but a little different than the average short crystallographic distance $R_{M-O}^{short} = 2.03$ Å. In short, first, the EXAFS results have revealed that the refined structure thus arises from averaging local structures for each type of ions, Fe or Co. Secondly, the $M - O$ bond anisotropy (*e.g.*, the difference in the bond distances between the long and short $M - O$ bond lengths) has been found to be in accordance with the crystallographic data.

The $Co$ addition gives rise to the distortions of the coordination octahedra as mentioned above. Namely, the $Co$ addition is accompanied by shifts in the oxygen positions and causes the increase in the bond lengths along the $c$ −axis (rhombic distortions) and decrease in the perpendicular direction (tetragonal ones), as seen from the normal coordinates $Q_2$ and $Q_3$. The largest distortion is a compression perpendicular to the $c$ −axis as seen from the planar and the axial average radii. The observed increase in $Q_3$ for M2 site can assign to FeO$_6$ octahedra with short axial bonds ~1.97 Å. As $Co$ content increases, the rhombic distortions increase for the M1 site. This is consistent with two extra electrons occupying a $d_{xy}$ orbital with lobes oriented parallel to the $c$ −axis. The distortions lift the $t_{2g}$ orbitals' degeneracy resulting in the energy gain associated with the Jahn-Teller effect. The $Co^{2+}$ ions push the $Fe^{3+}$ ones from the M1 site $(V_{zz} = 0.104\ e \cdot Å^3)$ to the M2 one $(V_{zz} = 0.087\ e \cdot Å^3)$, for which the $t_{2g}$ orbitals splitting should be smaller. In the studied warwickites, the $Fe^{3+}$ ions have the $S = 5/2$ configuration corresponding to a half-filled $d$ −orbital with $L = 0$ ($S$ state). Such a state is generally associated with a small magnetic anisotropy due to the absence of spin-orbit coupling. On the other hand, the $Co^{2+}$ ions ($3d^7$ electron configuration, $S = 3/2$, $L = 3$) being at the low-symmetry crystalline field of the coordinated octahedron display a strong coupling to the lattice. Taking into account the striking magnetic anisotropy properties of the parent ludwigites [38, 39], we expect that the $Co$-addition can induce the onset of magnetic crystalline anisotropy in the warwickites under study as well.

## 4. Conclusions

In the present work, a careful study of the electronic state, long-range, and local structural properties of three-component warwickite system $Mg_{1-x}Co_xFeBO_4$ has been carried out. XANES/EXAFS spectra analysis is supported by X-ray crystallographic data to build up a wide view of the electronic state and local structure of metal ions.

The X-ray diffraction and electron microanalysis data have shown that $Co$ and $Mg$ ions enter in the divalent states, and $Fe$ ions in the trivalent state. There is atomic disorder, since both crystallographic positions M1 and M2 are occupied by the mixture of $Mg$, $Co$, and $Fe$ atoms. The $Fe^{3+}$ ions occupy metal positions with a clearly pronounced preference; indeed, $Fe$ ions prefer to occupy the inner positions M1 in $Mg-Fe$ and $Mg-Co-Fe$ warwickites, and the outer positions M2 in the end member $Co-Fe$ warwickite, which has the shortest metal-oxygen distances. Low-symmetric Jahn-Teller-like distortions exist for both coordinated octahedra as can be seen from the normal coordinates calculations. As the $Co$ content increases, the structural distortions become more pronounced. An alternation of the octahedra principal axes within the ribbon was found.

The XRD data yield the average M - O distances over the two metal Fe and Co, so the increasing distortion cannot be assigned either to Fe or Co atoms. In spite of this, $V_{zz}$ calculated with those data increases with Co content and shows that Co atoms plays an important role in the structure modification. The effect of Co substitution is clarified by means of the element selective techniques XANES and EXAFS data. The average interatomic distances $Fe$ - $O$ and $Co$ - $O$ provide direct evidence for trivalent and divalent states of iron and cobalt ions, respectively. The electronic states of the Fe and Co are not affected by the substitution. The substituted Co ions have two roles: one is pushing the Fe atom from M1 site to M2 site, which is expressed in the reduction of the M2 site volume, and the other is an increase in the local distortions in $FeO_6$ and $CoO_6$ octahedra. In addition to XRD data, EXAFS analysis revealed that the Fe-O short bonds are smaller than the Co-O ones. So, the tetragonal distortions are expected to be greater for $FeO_6$ octahedra than for $CoO_6$ ones. At the same time the substitution of $Fe^{3+}$ ions by $Co^{2+}$ ions at the M1 site induces an increase in the rhombic distortion of $M1O_6$ octahedra for which the $t_{2g}$ orbitals splitting is larger.

Finally, we may infer that the local structure distortions around the magnetic $Fe$ and $Co$ atoms could play an important role in magnetic properties and in magnetic crystalline anisotropy, especially. The magnetic properties of the $Mg_{1-x}Co_xFeBO_4$ warwickite system will a matter of future study.

**Acknowledgments** This work has been financed by the MECOM Project MAT11/23791, and DGA IMANA project E-34, tCouncil for Grants of the President of the Russian Federation (project nos. NSh-2886.2014.2, SP-938.2015.5)., Russian Foundation for Basic Research (project nos. 13-02-00958-a, 13-02-00358-a, 14-02-31051-mol-a), The work of one of coauthors (M.S.P.) was supported by the grant of KSAI „Krasnoyarsk regional fund of supporting scientific and technological activities" and by the program of Foundation for Promotion of Small Enterprises in Science and Technology ("UMNIK" program).